\documentclass[letterpaper,preprintnumbers,amsmath,amssymb,nofootinbib,onecolumn,notitlepage,10pt]{revtex4-1}
\usepackage{graphicx}
\usepackage{dcolumn}
\usepackage{bm}
\usepackage[usenames,dvipsnames]{color}
\usepackage{amsmath}	
\usepackage{amssymb}	
\usepackage{multirow}
\usepackage{gensymb}
\usepackage[colorlinks=true,citecolor=blue,linkcolor=blue]{hyperref}
\include{jdefs}

\newcommand{\pstevflux}[1]{{$\rm{TeV^{-1}cm^{-2}s^{-1}}$}}

\begin{document}

\title{On the Neutrino Flares from the Direction of TXS 0506+056}

\author{Francis Halzen}
\email{francis.halzen@icecube.wisc.edu}
\author{Ali Kheirandish}
\email{ali.kheirandish@icecube.wisc.edu}
\author{Thomas Weisgarber}
\email{weisgarber@wisc.edu}
\affiliation{Department of Physics \& Wisconsin IceCube Particle Astrophysics Center, University of Wisconsin, Madison, WI 53706, USA}
\author{Scott P. Wakely}
\email{wakely@ulysses.uchicago.edu}
\affiliation{Enrico Fermi Institute \& Department of Physics, University of Chicago, Chicago, IL 60637, USA\\}

\begin{abstract}
For the first time since the discovery of high-energy cosmic neutrinos by IceCube, a multimessenger campaign identified a distant gamma ray blazar, TXS 0506+056, as the source of a high-energy neutrino. The extraordinary brightness of the blazar despite its distance suggests that it may belong to a special class of sources that produce cosmic rays. Moreover, over the last 10 years of data, the high-energy neutrino flux from the source is dominated by a previous neutrino flare in 2014, which implies that flaring sources strongly contribute to the cosmic ray flux. We investigate the contribution of this subclass of flaring blazars to the high-energy neutrino flux and examine its connection to the very high energy cosmic ray observations. We also study the high energy gamma ray emission accompanying the neutrino flare and show that the sources must be more efficient neutrino than gamma ray emitters. This conclusion is supported by the gamma-ray observations during the 2014 neutrino flare.
\end{abstract}
\maketitle

\section{Introduction}

The rationale for multimessenger astronomy is to search for the sources of cosmic rays by observation of high-energy neutrinos and gamma rays from pionic decays. High-energy cosmic neutrinos only originate in environments where protons are accelerated to produce pions and other particles that decay into neutrinos. 

The era of multimessenger astronomy with high energy neutrinos has begun with the discovery of cosmic neutrinos in 2013 \cite{Aartsen:2013jdh, Aartsen:2014gkd}. The continuous observation of the high-energy cosmic neutrino flux with contained events \cite{Kopper:2015vzf, Aartsen:2017mau} and throughgoing muons \cite{Aartsen:2015rwa, Aartsen2016a, Aartsen:2017mau} has shown possible features in the spectrum of high-energy cosmic neutrinos; however, until recently, it had not revealed the origin of the IceCube neutrinos \cite{Aartsen:2016oji, Aartsen:2017kru}. With the advantage of temporal coincidence, a flaring gamma ray blazar, TXS 0506+056, was identified in follow up observations as the source of a high-energy neutrino detected on September 22, 2017 \cite{IceCube:2018dnn}. Knowing where to look, the IceCube collaboration performed a search in the archival data obtained over a decade of detector operation, finding an excess of neutrinos in a flare that lasted $\sim$ 110 days in 2014. The single flare dominates the flux of the neutrinos from the direction of TXS 0506+056 over the 10-year period of observations \cite{IceCube:2018cha}. 

It has already been recognized that the most surprising property of the observed cosmic neutrino flux is its magnitude.
An important message emerging from the IceCube measurements of the high-energy cosmic neutrino flux is the prominent and surprisingly important role of protons relative to electrons in the nonthermal universe. The matching energy densities of the extragalactic gamma-ray flux detected by the Fermi Large Area Telescope (Fermi-LAT) and the high-energy neutrino flux measured by IceCube suggest a common origin. Moreover, the high intensity of the neutrino flux below 100~TeV in comparison to the Fermi data might indicate that these sources are even more efficient neutrino than gamma-ray sources~\citep{Murase:2015xka,Bechtol:2015uqb}. Interestingly, the common energy density of photons and neutrinos is also comparable to that of the ultra-high-energy extragalactic cosmic rays (above $10^{9}$~GeV). Unless accidental, this indicates a common origin and illustrates the potential of multimessenger studies. 

In this paper, we focus on the 2014 neutrino burst identified in the archival data and speculate on what that observation means in terms of the total diffuse neutrino flux observed in IceCube. Guided by the very large flux and luminosity produced by TXS 00506+056 in the 2014 flare, we study whether a subclass of blazars can explain the diffuse neutrino flux observed by IceCube. Given the similar energy in cosmic rays and neutrinos, we use the energy content of the very high-energy cosmic rays to understand the workings of this subclass of sources. 

We will show that a subset of about 5\% of all blazars, bursting once in 10 years at the level of TXS 0506+056 in 2014, can accommodate the diffuse cosmic neutrino flux observed by IceCube. The connection to the energy content of the very high-energy cosmic rays illustrates that the sources have a high efficiency, i.e., the sources are very efficient in producing neutrinos and, as a consequence, opaque to the accompanying high energy gamma rays. 

 A key question about the 2014 neutrino burst is: where are the high-energy gamma rays that should accompany the high-energy neutrinos? The gamma-ray flux measurement from the Fermi-LAT satellite is the focus of our study.
We incorporate the flux measurement during the neutrino flare in 2014 by \cite{Garrappa:2018}. Fermi data hint at a potential hardening of the spectrum \cite{Padovani:2018acg}. However, the slope change is not very significant \cite{Garrappa:2018}.
Using these measurements, we show that a consistent picture emerges when the source opacity creates a gamma-ray cascade at the source, followed by cascading to lower energies on the extragalactic background light (EBL) as the gamma rays propagate out of the source to Earth.

\section{Time-Dependent Neutrino Emission from TXS 0506+056}

In a multimessenger campaign, TXS 0506+056 was identified as a likely source of a $290$\,TeV neutrino observed in IceCube on September 2017 \cite{IceCube:2018dnn}. The identification of the source led to an archival study of the time-dependent and time-integrated neutrino emission from the data collected in 9.5 years of IceCube operation \cite{IceCube:2018cha}. The time-dependent analysis revealed in 2014 a large burst of $13\pm5$ neutrinos over a period of 110 days. This observation yields a significance of 3.5 $\sigma$, rejecting a background explanation for the events, and dominates the observed flux from the source over 10 years of observation.

TXS 0506+056 is an intermediate synchrotron-peaked BL Lac at a redshift of 0.34 \cite{Paiano:2018qeq}. During the 2014 burst, the neutrino flux was $1.6\times10^{-15}$\,\pstevflux  \, at 100 TeV \cite{IceCube:2018cha}. The time-averaged flux of neutrinos from the source over 9.5 years of IceCube observations was $0.8 \times 10^{-16} \mathrm {TeV}^{-1} \mathrm {cm} ^ {-2} \mathrm {s}^{-1}$ \cite{IceCube:2018cha}. 

TXS 0506+056, despite its greater distance, outshines nearby blazars, suggesting a special class of sources that accelerate protons and produce very high-energy gamma rays and neutrinos. The isotropic neutrino luminosity of TXS 0506+056 during the 2014 burst is $1.2\times10^{47}\rm{erg/s}$. We will use this luminosity in Section \ref{flaresec} to calculate the contribution of the subclass of sources to the total flux of neutrinos. Note that TXS 0506+056 had already been identified as a unique blazar in EGRET observations, preeminent for observation of two photons with energies above 40 GeV \citep{Dingus:2001hz}.
 Despite its large redshift, it is one of the brightest sources in Fermi catalogs \cite{Garrappa:2018}. 

\section{The Qualitative Features of the High-Energy Cosmic Neutrino Flux from the Flaring Blazar}\label{nufluxsec}

High-energy cosmic neutrinos are produced at astrophysical beam dumps where accelerated protons encounter sufficient target matter or radiation density to produce pions and other particles that decay to neutrinos. For blazars, one routinely focuses on the $p\gamma$ interactions of very high-energy protons with abundant target photons to produce charged and neutral pions. In order to produce pions, the proton energy must exceed the threshold for producing the $\Delta$-resonance. The neutrinos leave the site of production while pionic gamma rays may be absorbed or cascade to lower energies if the source is fully or partially opaque. The maximum energy with which gamma rays leave the source depends on the energy of the target photons and the Lorentz factor of the jet. After emission, the gamma rays are further absorbed via pair production on the EBL, further reducing their flux. It is also possible to obtain an increase in the flux at lower energies due to inverse Compton scattering of the produced pairs on the CMB.

The relation between the pionic gamma-ray and neutrino production rates is:
\begin{equation}\label{eq:GAMMAtoNU}
\frac{1}{3}\sum_{\alpha}E^2_\nu Q_{\nu_\alpha}(E_\nu) \simeq \frac{K_\pi}{4}\left[E^2_\gamma Q_\gamma(E_\gamma)\right]_{E_\gamma = 2E_\nu}\,.
\end{equation}
The prefactor $1/4$ accounts for the energy ratio $\langle E_\nu\rangle/\langle E_\gamma\rangle\simeq 1/2$ and the two gamma rays produced in the neutral pion decay \cite{Ahlers:2018fkn}. Note that this relation relates pionic neutrinos and gamma rays without any reference to the cosmic ray beam; it simply reflects the fact that a neutral pion produces two gamma rays for every charged pion producing a $\nu_\mu +\bar\nu_\mu$ pair, which cannot be separated by current experiments. The factor $K_\pi$ accounts for the ratio of charged to neutral pions. $K_\pi\simeq2$ for cosmic-ray interactions with gas ($pp$) and $K_\pi\simeq1$ for interactions with photons ($p\gamma$) \citep{Ahlers:2018fkn}.

\section{Diffuse Cosmic Neutrino Flux and Flaring Blazars}\label{flaresec}

The extraordinary detection of more than a dozen cosmic neutrinos in the 2014 flare despite the 0.34 redshift of the source suggests that TXS 0506+056 belongs to a special class of sources that produce cosmic rays. The single neutrino flare dominates the flux of the source over the 9.5 years of archival IceCube data, leaving IC 170922A as a less luminous second flare in the sample. 

In this section we try to answer three major questions: what is special about this source, can a subclass of blazars with similar characteristics accommodate the diffuse flux observed by IceCube, and how do these sources contribute to the flux of the very high-energy cosmic rays?

In order to calculate the flux of high-energy neutrinos from a population of sources, we follow \citep{Halzen:2002pg} and relate the diffuse neutrino flux to the injection rate of cosmic rays and their efficiency to produce neutrinos in the source. For a class of sources with density $\rho$ and neutrino luminosity $L_\nu$, the all-sky neutrino flux is
\begin{eqnarray}
\sum_{\alpha} E_\nu^2 \frac{d N_\nu}{dE_\nu} = \frac{1}{4\pi} \frac{c}{H_0} \xi_z L_\nu \rho,
\end{eqnarray}
where $\xi_z$ is a factor of order unity that parametrizes the integration over the redshift evolution of the sources. 
The relation can be adapted to a fraction $\mathcal{F}$ of sources with episodic emission of flares of duration $\Delta t$ over a total observation time $T$:
\begin{eqnarray}
\sum_{\alpha} E_\nu^2 \frac{d N_\nu}{dE_\nu} = \frac{1}{4\pi} \frac{c}{H_0} \xi_z L_\nu \rho \mathcal{F} \frac{\Delta t}{T} \,.
\end{eqnarray}
Applying this relation to the 2014 TXS 0506+056 neutrino flare and the density of BL Lacs \citep{Mertsch:2016hcd} yields
\begin{eqnarray}\label{diffuse_flux}
\begin{aligned}
3\times10^{-11}\, {\rm{TeV cm^{-2} s^{-1} sr^{-1}}} = &\frac{ \mathcal{F}}{4\pi} \frac{c}{H_0} \bigg(\frac{\xi_z}{0.7 \rm{}} \bigg) \bigg(\frac{L_\nu}{1.2\times10^{47}\, \rm{erg/s}} \bigg)\\ &\bigg(\frac{\rho}{1.5\times10^{-8}\, \rm{Mpc^{-3}}} \bigg) \bigg(\frac{\Delta t}{110 \,{\rm d}} \frac{10 {\, \rm yr}}{T}\bigg)\,, 
\end{aligned}
\end{eqnarray}
a relation which is satisfied for $\mathcal{F}\sim0.05$. This means that a special class of blazars that undergo $\sim110$-day duration flares like TXS 0506+056 once every 10 years accommodates the observed diffuse flux of high-energy cosmic neutrinos. 

With the matching energetics of the cosmic neutrinos and the very high-energy cosmic rays we can investigate the efficiency of the sources to produce high-energy cosmic neutrinos. The diffuse high-energy cosmic neutrino flux is related to the energy flux of the cosmic rays by
\begin{eqnarray}
\frac{1}{3}\sum_{\alpha} E_\nu^2 \frac{d N_\nu}{dE_\nu} \simeq \frac{c}{4 \pi}\,\bigg( \frac{1}{2}
(1-e^{-f_\pi})\, \xi_z t_H \frac{dE}{dt} \bigg)\,.
\end{eqnarray}
The cosmic rays' injection rate $dE/dt$ above $10^{16}$ eV is $(1-2) \times 10^{44}\,\rm erg$\, $\rm Mpc^{-3}\,yr^{-1}$ \citep{Ahlers:2012rz, Katz:2013ooa}. From Equation \ref{diffuse_flux} it follows that the energy densities match for a pion production efficiency of the neutrino source of $f_\pi \gtrsim 0.4$. This high efficiency requirement is consistent with the premise that a special class of efficient sources is responsible for producing the high-energy cosmic neutrino flux seen by IceCube. The sources must contain sufficient target density in photons, even possibly protons, to generate the large value of $f_\pi$. It is clear that the emission of flares producing the large number of cosmic neutrinos detected in the 2014 burst must correspond to major accretion events onto the black hole lasting a few months. The pionic photons will lose energy in the source and the neutrino emission is not accompanied by a flare as was the case for the 2017 event. The Fermi data are consistent with the scenario proposed; they reveal photons with energies of tens of GeV, but no flaring activity.

A key question is whether the neutrino and gamma ray spectra for the 2014 neutrino burst from TXS 0506+056 satisfy the multimessenger relationship introduced in Section \ref{nufluxsec}. With the low statistics of the high-energy gamma ray measurements during the burst period, the energetics represents a more robust measure for evaluating the connection, especially because the source is opaque to high-energy gamma rays as indicated by the large value of $f_\pi$. The pionic gamma rays will lose energy inside the source before cascading in the EBL. Here, we will specifically use the connection between the $p\gamma$ and $\gamma\gamma$ opacities introduced in \cite{Murase:2015xka}:
\begin{equation}
\tau _ { \gamma \gamma } \approx \frac { \eta _ { \gamma \gamma } \sigma _ { \gamma \gamma } } { \eta _ { p \gamma } \hat { \sigma } _ { p \gamma } } f _ { \pi },
\end{equation}
where $\hat {\sigma} _ {p\gamma} \sim 0.7 \times 10 ^{-28} \mathrm {cm}^{2}$ and $\sigma_{\gamma\gamma} \simeq 6.65 \times 10 ^ { - 25 } \mathrm{cm}^2$. We assume $\eta _ { \gamma \gamma } \sim 0.1$ and $\eta _ { p \gamma } \simeq 1$. Considering $f_\pi \gtrsim 0.4$ yields $\tau _ { \gamma \gamma } \simeq \mathcal{O}(100)$. This illustrates a very high opacity for the source, making it impossible for the very high-energy pionic gamma rays, with energies similar to those of the neutrinos, to leave the source.

\section{High-Energy Gamma ray emission accompanying the 2014 Flare}

The gamma-ray flux initiated by neutral pion decays associated with the neutrino burst is reprocessed to lower energies because of the opacity of the source and is unlikely to be directly observable.
To investigate the consistency of the overall picture, we adopt a phenomenological model of the high-energy gamma rays emerging from the source.
The spectrum after internal reprocessing is given by
\begin{equation}
\frac{dN}{dE}=AE^{-2}\mathrm{e}^{-E_L/E-E/E_H},
\label{eqn:source-output}
\end{equation}
where $E_L$ and $E_H$ are low- and high-energy cutoffs, and $A$ is chosen to match the lower bound on the total power emitted in gamma rays between 30 TeV and 3 PeV consistent with the IceCube neutrino observations.
Thus, we assume that there are no internal losses during reprocessing and that total power in gamma rays is conserved.
We further make the approximation that this gamma-ray spectrum is constant over the 110 days of the neutrino burst.

The spectrum given by Equation \ref{eqn:source-output} suffers attenuation via pair production interactions with the EBL.
In principle, the electron-positron pairs produced in these interactions can upscatter CMB photons into the high-energy regime, producing a cascaded component in addition to the attenuated, direct emission~\citep{Aharonian:1994ix,Plaga:1995hv}.
However, this cascade will only be observable if the intergalactic magnetic field (IGMF) is weak enough that the pairs are not deflected too much before cooling on the CMB.
We model the effects of the EBL and IGMF using a particle-tracking simulation similar to the one presented in~\citep{Arlen:2014if} and used in~\citep{Archambault:2017cn}.
This simulation accounts for the full relativistic cross sections of the pair production and inverse Compson scattering processes and allows for arbitrary EBL and IGMF evolution with redshift.
Our model for the IGMF assumes a coherence length of 1 Mpc and uses the methods presented in~\citep{Giacalone:1999ci} to achieve a smoothly varying randomly oriented field with strength $B_\mathrm{IGMF}$.
We adopt the EBL model of~\citep{Gilmore:2012il}, and we further assume that interactions with the CMB are the dominant energy loss mechanism for the pairs, although this is under debate, see, e.g. ~\citep{Broderick:2012da,Sironi:2014bj,Menzler:2015jk,Chang:2016gh}.
In running the simulation to model the observed photon spectrum during the neutrino burst, we remove any gamma rays that arrive with time delays larger than 110 days.

\begin{figure}
\begin{center}
\includegraphics[width=0.75\textwidth]{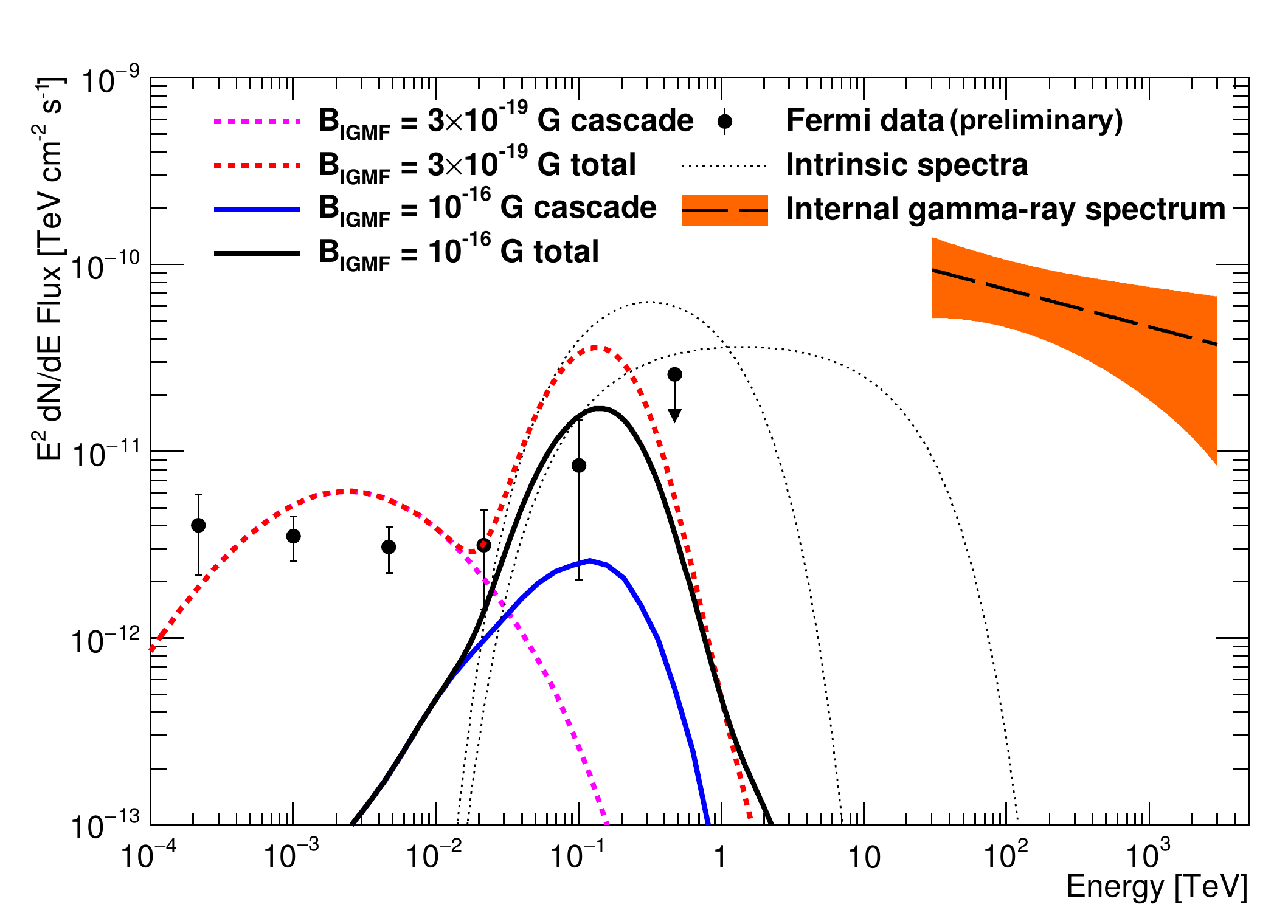}
\caption{
Observed cascade and total spectra for our model with $E_H=20$ TeV and $B_\mathrm{IGMF}=10^{-16}$ G (thick solid lines), and for $E_H=1$ TeV and $B_\mathrm{IGMF}=3\times10^{-19}$ G (thick dashed lines), along with the Fermi observations and internal pionic gamma rays associated with the 2014 neutrino outburst.
The low-energy cutoff is $E_L=100$ GeV for both cases.
The intrinsic spectra (thin dashed lines) are given by Equation~\ref{eqn:source-output} and represent emission from the source after reprocessing but before attenuation on the EBL.
}
\label{fig:spectra}
\end{center}
\end{figure}

We find that a large value of the low-energy cutoff, $E_L\gtrsim100$ GeV, is required to produce consistency with the Fermi observations during the 2014 neutrino outburst.
Figure~\ref{fig:spectra} shows our results for two cases.
For the first, we assume $E_L=100$ GeV, $E_H=1$ TeV, and $B_\mathrm{IGMF}=3\times10^{-19}$ G.
The cascade emission in this case reproduces the Fermi data well.
However, an IGMF this weak is in tension with recent results from the Fermi collaboration~\citep{Ackermann:2018cc} and may not be viable.
In our second case, we leave the low-energy cutoff fixed and assume $B_\mathrm{IGMF}=10^{-16}$ G, compatible with the Fermi limits.
This allows us to raise the high energy cutoff to $E_H=20$ TeV because the cascade is strongly suppressed at lower energies.
In this case, the combined cascade and direct emission, as shown in Figure~\ref{fig:spectra}, can accommodate the apparent hardening of the Fermi spectrum due to its contribution in the energy range above $\sim$10 GeV, although the Fermi emission at lower energies would have to be produced by some other process.
Although we only show the result for $E_H=20$ TeV in Figure~\ref{fig:spectra}, we find that for $B_\mathrm{IGMF}=10^{-16}$ G the direct emission for any value of $E_H$ between 500 GeV and 20 TeV produces an acceptable description of the Fermi data above $\sim$10 GeV.

\section{Discussion \& Conclusion}

The evidence for neutrino emission from TXS 0506+056 has set a tipping point in the search for the sources of high-energy cosmic neutrinos. Getting all the elements of this puzzle to fit together is not easy, but they suggest that the blazar may contain important clues on the origin of cosmic neutrinos and cosmic rays. This breakthrough is just the beginning and raises intriguing questions. Here we tried to address a few of these questions: What is special about this source? Can a subclass of blazars to which it belongs accommodate the diffuse flux observed by IceCube? Are these also the sources of all high-energy cosmic rays or only of some?  

We explored the contribution of a subclass of blazars with characteristics similar to those of TXS 0506+056 during the 2014 burst and what that means for total flux measurements in IceCube. The class of such neutrino-flaring sources represents 5\% of the sources. 

The high level of neutrino flux from TXS 0506+056 requires very efficient neutrino production at the source. We examined the efficiency by connecting the total neutrino flux to the energy content of the extragalactic cosmic rays, incorporating the common energy in high-energy cosmic neutrinos and the extragalactic cosmic rays. We find a high photohadronic efficiency that indicates high gamma-ray opacity of the sources. This means that the sources emit neutrinos more efficiently than they emit gamma rays.

We further examined the gamma ray emission during the dominant neutrino flare in 2014. Our results show that the absorption and interactions intrinsic to the source, followed by the interaction with EBL, will result in a gamma ray flux consistent with the Fermi observations. A gamma ray flare is not expected when the source is a highly efficient neutrino emitter. We note that, since we do not consider internal losses at the source, our estimated gamma ray flux represents an upper limit on the expected gamma ray emission accompanying the neutrinos in 2014. 
The TXS 0506+056 neutrino emission over the last 10 years is dominated by the single flare in 2014. If this is characteristic of the subclass of sources that it belongs to, identifying additional sources will be difficult unless more and larger neutrino telescopes yield more frequent and higher statistics neutrino alerts. 

\section{Acknowledgments}
We thank Nahee Park, Ibrahim Safa, Maria Petropoulou, and Simone Garrappa for comments and discussions. We especially thank Jay Gallagher for his useful comments during this study. FH and AK are supported in part by NSF under grants PLR-1600823 and PHY-1607644 and by the University of Wisconsin Research Committee with funds granted by the Wisconsin Alumni Research Foundation. TW is supported by NSF under the grant PHY-1707635.

\bibliographystyle{apsrev4-1}
\bibliography{bib}

\end{document}